\documentstyle[12pt]{article}
\begin{document}
\topmargin 0pt
\oddsidemargin -3.5mm
\headheight 0pt
\topskip 0mm
\addtolength{\baselineskip}{0.20\baselineskip}
\begin{flushright}
SOGANG-HEP $202/95$
\end{flushright}
\vspace{1.0cm}
\begin{center}
{\Large \bf  Investigation on the Tachyonic Neutrino }
\end{center}
\vspace{0.0cm}
\begin{center}
{Mu-In Park$^{a}$ and Young-Jai Park$^{b}$}\\
\vspace{0.0cm}
{Department of Physics and Basic Science Research Institute }\\
{Sogang University, C.P.O.Box 1142, Seoul 100-611, Korea}\\
\vspace{1.0cm}
{\large \bf ABSTRACT}
\end{center}
According to the experimental data, it is still controversial whether the
neutrinos, especially the electron-neutrino and muon-neutrino, can be
considered
as the fermionic spinorial tachyons, and there is still no reliable report on
the existence of the right-handed neutrinos. In this letter, we show that the
neutrinos with the single handedness can not be the
tachyons, but only those of the both handedness can be. Several implications of
this result are discussed.

\vspace{7.0cm}
\begin{flushleft}
September 1995 \\
$^{a}$ Electronic address: mipark@physics.sogang.ac.kr \\
$^{b}$ Electronic address: yjpark@physics.sogang.ac.kr \\
\end{flushleft}

\newpage

According to recent experimental data [1], it is still controversial
whether the neutrinos, especially the electron-neutrino and
muon-neutrino, can be considered as the tachyons [2], which is the hypothetical
objects moving faster than light in vacuum
\footnote[1]{
The pioneering works on the usual tachyon theory were given by Ref. [2].
Later development were
not deviated much far from the lines of these papers. However, it has
been pointed out recently that these formulations are incomplete even at
the classical (non-quantum theoretical) level and the
theory was reformulated by authors, On the
Foundation of the Relativistic Dynamics with the Tachyon, Sogang
Univ. Report No. SOGANG-HEP 197/95, hep-th/9506082. According to the our
formulation, the rest mass of the tachyon is not anymore Lorentz scalar but the
sign may be changed under the Lorentz transformation depending on it's velocity
for consistency. But even in this
formulation, the mass squared ${m^{2}}$ is still Lorentz invariant. Hence, in
our
interesting wave equation in this letter, the linear wave equation, this
unusual
property of the mass may have a role for the covariance of the wave equation.
But we will not quote here this new formulation because it is sufficient to
use,
to derive our result, only the fact that mass squared $m^{2}$ for the tachyon
is
negative valued Lorentz scalar, which is the same for the both old and new
formulations.},
i.e., fermionic
spinorial tachyons. Furthermore, there is still no reliable report on the
existence of the right-handed neutrinos.

In this letter, we firstly present
a proof that the single handedness particles can not be the
tachyons, but only  those of the both handedness can be.

Let us start by considering the tachyonic Dirac equation [3], which is Lorentz
covariant and represents the tachyonic particle in 4$\times$4 representation
in the physical 3-space and 1-time dimension as follows
\begin{eqnarray}
\left( i \gamma^{\mu} \partial_{\mu} -\lambda_{T} \right) \psi(x)~ = ~0,
\end{eqnarray}
where $\lambda_{T}$ is generally given by
\begin{eqnarray}
\lambda_{T}=i a I+b \gamma^{5}
\end{eqnarray}
with the real constants $a$ and $b$, and $\gamma^{5} \equiv i \gamma^{0}
\gamma^{1}
\gamma^{2} \gamma^{3}$ \footnote[2]{
The general wave equations for the spinorial bradyon (the object
moving slower than light) and luxon (the object moving with the velocity
of light) can be described similarly by $\lambda_{B}=c I +i d \gamma^{5}$
and $\lambda_{L}=f_{\pm} (I \pm \gamma^{5})$ for real numbers $c,~d$
and complex number $f_{\pm}$, respectively. Furthermore, note that we can use
the chiral transformation to transform the pseudoscalar or scalar part away.
But, in that case the physics described by the spinor is changed due to the
non-invariance of the theories under the transformation.}.
By way of the group theory, this finite-component
(here four-component) theory, of course,
involves a non-unitary representation of the Lorentz transformation. Here,
$\gamma-$matrices satisfies the usual
Clifford algebra $\{ \gamma^{\mu}, \gamma^{\nu} \}=2 g^{\mu \nu}$ in our
convention $g^{\mu \nu}=\mbox{diag}(1,-1,-1,-1)$. In this case the
mass squared $m^{2}$ of the spinor $\psi(x)$ is found to be
\begin{eqnarray}
m^{2}=(\gamma^{0} \lambda_{T})^{2}=-a^{2}-b^{2}~~<~~0
\end{eqnarray}
implying clearly the tachyonic movement, by comparing with the Klein-Gordon
equation
\begin{eqnarray*}
\left( \partial^{\mu} \partial_{\mu}+m^{2} \right) \psi(x)~ =~0.
\end{eqnarray*}
Moreover, in this wave equation the usual vector current
$J^{\mu}\equiv \bar{\psi} \gamma^{\mu} \psi $ is not conserved anymore for any
non-zero
$a$ and $b$, but the axial current $J^{\mu}_{5}\equiv \bar{\psi}
\gamma^{\mu} \gamma^{5} \psi $ can be conserved for the case of $a=0$
discarding the physically uninteresting case of $b=\infty$, which corresponds
to the infinitely massive immovable particle because of
\begin{eqnarray}
\partial_{\mu}J^{\mu}
& =&2 \left[ a \bar{\psi} \psi -i b \bar{\psi} \gamma^{5} \psi \right], \\
\partial_{\mu}J^{\mu}_{5}
& =&\frac{-ia}{b}   \bar{\psi}\gamma^{\mu}
\stackrel{\leftrightarrow}{\partial}_{\mu} \psi,
\end{eqnarray}
where $\stackrel{\leftrightarrow}{\partial}_{\mu}$ acts as
$F \stackrel{\leftrightarrow}{\partial}_{\mu} G=
F \partial_{\mu} G-(\partial_{\mu} F)G$ for some function $F$ and $G$.
The corresponding Lagrangian density, which is Hermitian or anti-Hermitian
depending
on the statistics \footnote[3]{
In general, this Hermicity or anti-Hermicity guarantee the
consistency of the Euler-Lagrange equations derived from the variation
of $\psi$ and $\psi^{*}$. But, the Hermicity is favored such that in
this case the Hermicity of the Hamiltonian is also guaranteed.}
is uniquely found to be (up to normalization constant)
\footnote[4]{
Including the bradyon and luxon cases, all cases are
described unifiedly by ${\cal L}_{\xi}=(i/2)\bar{\psi} \xi
\gamma^{\mu} \stackrel{\leftrightarrow}{\partial}_{\mu} \psi
- \bar{\psi} \xi \lambda \psi$
with $\xi=-\gamma^{5},~I$, and $(I \pm \gamma^{5} )$ for the cases of
tachyon, bradyon, and luxon, respectively.}
\begin{eqnarray}
 {\cal{L}}_{T}=-\frac{i}{2} \bar{\psi} \gamma^{5} \gamma^{\mu}
 \stackrel{\leftrightarrow}{\partial}_{\mu} \psi
 +  \bar{\psi} \gamma^{5} \lambda_{T} \psi.
\end{eqnarray}
Furthermore, the corresponding canonical Hamiltonian becomes
\begin{eqnarray}
H_{T}&=&\int d^{3}x \left( \Pi_{\psi} \dot{\psi}+\dot{ \psi}^{\dagger}
\Pi_{\psi^{*}}
-{\cal L}_{T} \right) \\ \nonumber
    &=&\int d^{3} x \psi^{\dagger} \gamma^{5} h_{T} \psi,
\end{eqnarray}
which is Hermitian or anti-Hermitian depending on the Hermicity of the
Lagrangian. Here, the canonical momenta are $\Pi_{\psi}\equiv
\frac{\partial {\cal L}}{\partial \dot{\psi}}|_{r}
=-(i/2)\bar{\psi} \gamma^{5} \gamma^{0}$ and $\Pi_{\psi ^{*}}\equiv
\frac{\partial {\cal L}}{\partial \dot{\psi}^{*}}|_{l}=-(i/2) \gamma^{5} \psi$,
where the subscripts $r$ and $l$ represent the right and left
derivatives, respectively, \footnote[5]{
The use of the different derivatives for $\Pi_{\psi}$ and
$\Pi_{\psi ^{*}}$  together with the unusual definition of the Hamiltonian (7)
(the formal sign of the second term is different with the usual one) are
devised in order that these can be defined without explicit consideration of
the exchange algebras.}
and $h_{T}$ is the one-particle Hamiltonian
\begin{eqnarray}
h_{T}=\vec{\alpha} \cdot \vec{ p} + i a \gamma^{0}+b \gamma^{0} \gamma^{5},
\end{eqnarray}
with $\vec{\alpha}=\gamma^{0} \vec{\gamma}$. Note that for the case of the
anti-Hermitian Hamiltonian the normalization constant of the Lagrangian density
${\cal L}_{T}$ should be adjusted such that the Hamiltonian is Hermitian. But,
in this letter we will preserve the normalization as the Lagrangian (6) since
this is not important problem in our analysis.

Now, in order to treat the handedness problem, let us explicitly consider the
chiral representation, i.e.,
\begin{eqnarray}
\psi =\left( \begin{array}{c}
             \psi_{L} \\
             \psi_{R} \\
            \end{array} \right) ,
{}~\gamma^{\mu} =\left(  \begin{array}{cc}
              0 &   \overline{\sigma}^{\mu} \\
              \sigma^{\mu} &     0          \\
               \end{array} \right),
{}~\gamma^{5} = \left( \begin{array}{cc}
               1 & 0 \\
               0 & -1  \\
               \end{array} \right),
\end{eqnarray}
where $\psi_{L}$ and $\psi_{R}$ are the two-component spinors that transform as
$({\bf \frac{1}{2}},{\bf 0})$ and $({\bf 0},{\bf \frac{1}{2}})$ representations
of the Lorentz group, respectively,
and $\sigma^{\mu} \equiv (1, \vec{\sigma} ), \overline{\sigma}^{\mu} \equiv
(1, - \vec{\sigma} )$. In this representation, Eq. (1) reduces to two sets of
equations
\begin{eqnarray}
i \overline{\sigma}^{\mu} \partial_{\mu} \psi_{R} -(i a +b) \psi_{L}~=~0, \\
i \sigma^{\mu} \partial_{\mu} \psi_{L} -(i a -b) \psi_{R}~=~0,
\end{eqnarray}
where both $\psi_{L}$ and $\psi_{R}$ are the tachyonic spinors having the same
mass squared $m^{2}=-a^{2}-b^{2}$ as that of $\psi$. Then, the
corresponding Lagrangian density and Hamiltonian become
\begin{eqnarray}
{\cal L}_{T} &=&\frac{i}{2} \psi^{\dagger}_{L}\sigma^{\mu}
     \stackrel{\leftrightarrow}{\partial}_{\mu} \psi_{L}
             - \frac{i}{2} \psi^{\dagger}_{R}\overline{\sigma}^{\mu}
             \stackrel{\leftrightarrow}{\partial}_{\mu} \psi_{R}
              +(ia +b) \psi^{\dagger}_{L} \psi_{R}
              +(-ia+b) \psi^{\dagger}_{R} \psi_{L}, \\
H_{T}& =&\int d^{3}x \left[
   -\frac{i}{2} \psi^{\dagger}_{L}\sigma^{i}
   \stackrel{\leftrightarrow}{\partial}_{i}\psi_{L}
              -\frac{i}{2} \psi^{\dagger}_{R}\sigma^{i}
              \stackrel{\leftrightarrow}{\partial}_{i}\psi_{R}
              -(ia +b) \psi^{\dagger}_{L} \psi_{R}
              -(-ia+b) \psi^{\dagger}_{R} \psi_{L}
            \right].
\end{eqnarray}
Moreover, the axial current density $J^{\mu}_{5}$, which is conserved for the
case of $a=0$, becomes
\begin{eqnarray}
J^{\mu}_{5}=\psi^{\dagger}_{L} \sigma^{\mu} \psi_{L}
-\psi^{\dagger}_{R}\overline{\sigma}^{\mu} \psi_{R}
\end{eqnarray}
explicitly showing the non positive-definiteness (more exactly the sign
indefiniteness) of $J^{0}_{5}=\psi^{\dagger}_{L}\psi_{L}
-\psi^{\dagger}_{R}\psi_{R}$ such that the usual probability interpretation
is questionable in this case. However, according to our usual experiences in
the
second quantization theory, this problem is not so serious one. In this case
it is well interpreted only if we can develop a theory with (lower) bounded
Hamiltonian
irrespective on the non-existence of the positive definite conserved current
density, of course, together with other fundamental principles like as the
microscopic causality and the Lorentz covariance. The bradyonic scalar,
spinor, and vector particles are the examples [4]. But, unfortunately this
scenario for
the tachyons can not be checked at present because there
are no known consistent second quantization rules for the spinorial tachyon.
Actually even for the tachyonic scalar, which will be the most simple case in
the tachyonic particles, the consistent quantization rule has not been known
so far [5]. However, we will show that especially
for single handedness spinorial particles we have a stringent situation for the
existence of the tachyons, i.e., the theory of the
tachyons with the single handedness like as the Majorana particles are not
consistent
even at the level of the first quantization without knowing the full situation
of the second quantized theory.

To prove this, let us consider the single handedness tachyons,
which is obtained directly from the both handedness theory by
reducing the handedness. On the other hand, we note that the most general
covariant reduction of
the handedness should be obtained, if the single-handed theory as well as the
both-handed theory are existed, by
\begin{eqnarray}
 \psi_{R}=- \alpha \sigma^{2} \psi^{*}_{L}
\end{eqnarray}
or
\begin{eqnarray}
 \psi_{L}= \beta \sigma^{2} \psi^{*}_{R}
\end{eqnarray}
since this is the most general relation connecting the two different handedness
spinors
$\psi_{R}$ and $\psi_{L}$ within the transformation theory of the spinor [6],
and hence the four-component spinor $\psi$ in Eq. (9)
becomes
\begin{eqnarray}
\psi =\left( \begin{array}{c}
             \psi_{L} \\
            -\alpha \sigma^{2} \psi^{*}_{L} \\ \end{array} \right)
\end{eqnarray}
or
\begin{eqnarray}
\psi =\left( \begin{array}{c}
             \beta \sigma^{2}\psi^{*}_{R} \\
            - \psi_{R}      \end{array} \right)
\end{eqnarray}
for the left-handedness or the right-handedness only theories with the
constant $\alpha$ and $\beta$, respectively. But, it is important to note
that the
${\it handedness-reduction}$, by it's means, should not change the physical
contents of each handedness spinor of the original theory except reducing the
handedness.

Now, for the application to the neutrinos we consider the only
left-handedness case (17). But, the conclusion is also the same for
the only right-handedness case (18). We first consider the reduction in the
wave equations, i.e., the reduction from the wave equations (10) and (11). By
putting spinor relation (15) into Eqs. (10) and (11), or equivalently (17) into
Eq. (1) we obtain
\begin{eqnarray}
i \alpha
\overline{\sigma}^{\mu}\sigma^{2}\partial_{\mu}\psi^{*}_{L}-(ia +b)
\psi_{L}~=~0, \\
i \sigma^{\mu}\partial_{\mu}\psi_{L}+\alpha(ia -b)\sigma^{2}\psi^{*}_{L}=0.
\end{eqnarray}
By inspection, it is easy to expect that Eqs. (19) and (20)
would be the complex conjugations of each other if these equations are
consistent.
However, surprisingly this is not the case. To see this, we apply the complex
conjugation to Eq. (19), and use the identity of the Pauli's spin matrices
\begin{eqnarray*}
\overline{\sigma}^{\mu}\sigma^{2}=\sigma^{2}\widetilde{\sigma^{\mu}}
\end{eqnarray*}
with the transposed matrices $\widetilde{\sigma^{\mu}}$. Then we find that
Eq. (19) can be written as follows
\begin{eqnarray}
i \sigma^{\mu}\partial_{\mu}\psi_{L}
  -\frac{1}{\alpha^{*}}(ia -b)\sigma^{2}\psi^{*}_{L}=0,
\end{eqnarray}
which should be equal to Eq. (20) for consistency.
But, this equation is equal to Eq. (20)
only if
\begin{eqnarray}
\alpha \alpha^{*} =-1
\end{eqnarray}
is satisfied since $a$ or $b$ is non-zero for the tachyons and
$\sigma^{\mu}\partial_{\mu}\psi_{L}\neq 0$ in general.
But, note that this has no solution within the complex number \footnote[6]{
We may introduce an hypothetical number $\alpha$ having
the property $\alpha \alpha^{*}~<~0$ by enlarging the set of numbers
in mathematics together with the spinor having the property of the
hypothetical number for each component. But, in this letter we only confine
ourselves to the usual number theory for simplicity. If this possibility is
considered, our
conclusion will be drastically changed. See Footnote 10 for this problem.}.
Hence,
the single handedness spinorial tachyon wave equation, which is reduced
from the both handedness spinorial tachyon wave equation, is inconsistent
for any non-zero mass and any complex-number $\alpha$. Furthermore, we note
that this
inconsistency can not be attributed to the matter of the handedness reduction
method due to it's general form, and hence it should be attributed to the
matter
of the spinorial tachyon wave equation itself. In other words, although we have
shown that the inconsistency of only the reduced single-handed tachyons from
the both-handed tachyons, this inconsistency is actually a genuine property of
the
single-handed tachyons themselves irrespective of the handedness reduction
method. Of course, for the zero-mass case of $a=b=0$
the consistency is trivially satisfied as is
well-known, for example, in the Weyl equation. Furthermore, note that this
result is derived without restricting to any statistics. Usually the
statistics of the particles is not determined by their wave equation,
but by the consistency of the second quantized theory with many
fundamental principles like as the ones mentioned previously. However, in many
cases \footnote[7]{
The real scalar field, complex scalar field when
decomposed into two real scalar fields, massless and massive vector particles
are the cases.
In these cases the classical Lagrangian and Hamiltonian become vanishing for
wrong
statistics.} the statistical nature of the fields can be also predicted from
the classical Lagrangians or Hamiltonians.

We now examine how the inconsistency at the level of equations of motion is
transferred to the Lagrangian and Hamiltonian by explicitly considering
the exchange algebras of the fields, which will be related to the statistics of
the fields after the second quantization. Moreover, since all the results from
the Hamiltonian analysis
can be also obtained in the Lagrangian analysis, we only consider here the
latter
analysis to avoid the duplication. To this end, let us
replace $\psi_{R}$ with $\psi_{L}$ by Eq. (15) in the Lagrangian
(12). Then, the Lagrangian reduces to
\begin{eqnarray}
{\cal L}_{T}=\frac{i}{2}\psi^{\dagger}_{L}\sigma^{\mu}
 \stackrel{\leftrightarrow}{\partial}_{\mu}\psi_{L}
-\frac{i}{2}|\alpha|^{2} \widetilde{\psi_{L}} \widetilde{\sigma^{\mu}}
     \stackrel{\leftrightarrow}{\partial}_{\mu}
 \psi^{*}_{L}-\alpha(ia+b) \psi^{\dagger}_{L} \sigma^{2}
 \psi^{*}_{L} - \alpha^{*} (-ia+b)\widetilde{\psi_{L}}\sigma^{2} \psi_{L}
\end{eqnarray}
without assuming the exchange algebras of the spinors. But, since the complete
analysis of the Lagrangian is possible only after explicit consideration of the
exchange algebras of the fields, we consider here the two typical cases,
i.e., anti-commuting and commuting fields, which will be corresponded to the
fermion and boson statistics after the second quantization, respectively. More
general
exchange algebras might be introduced, but that is not essential for our
consideration.

Firstly, let us consider the case of the anti-commuting fields, i.e.,
\begin{eqnarray}
\psi^{\dagger}_{L}(x)\psi_{L}(y)&=&-\psi_{L}(y)\psi^{\dagger}_{L}(x),
\nonumber \\
\psi_{L}(x)\psi_{L}(y)&=&-\psi_{L}(y) \psi_{L}(x).
\end{eqnarray}
After the second quantization these relations would be centrally deformed with
the operator-valued $\psi_{L}$ and $\psi_{L}^{\dagger}$.
Now, with the algebra (24), the Lagrangian (23)
reduces to
\begin{eqnarray}
{\cal L}_{T}=-\frac{i}{2}(|\alpha|^{2}-1)\psi^{\dagger}_{L}\sigma^{\mu}
\stackrel{\leftrightarrow}{\partial}_{\mu} \psi_{L}
-\alpha
(ia+b) \psi^{\dagger}_{L}\sigma^{2}\psi^{*}_{L}-\alpha^{*}(-ia+b)
\widetilde{\psi_{L}}\sigma^{2} \psi_{L}.
\end{eqnarray}
For the case of $|\alpha|=1$, which is the only case of the
consistent reduction of the bradyon, which will be shown later, the Lagrangian
has no kinetic terms, i.e.,
\begin{eqnarray}
{\cal L}_{T}=-\alpha
(ia+b) \psi^{\dagger}_{L}\sigma^{2}\psi^{*}_{L}-\alpha^{*}(-ia+b)
\widetilde{\psi_{L}}\sigma^{2} \psi_{L}.
\end{eqnarray}
Hence the handedness reduction of the relation (15) is failed in this case
because the handedness reduction changes the physical content of the spinor
$\psi_{L}$, i.e., it's mass can be considered to become infinitely large upon
the reduction even when the mass of the original both-handed Lagrangian (12)
is finite. However, note that this result is consistent with the
equation of motion analysis. In other words, for the case of $|\alpha|=1$
Eq. (21) equivalent to Eq. (19), becomes
\begin{eqnarray}
i \sigma^{\mu} \partial_{\mu} \psi_{L}-\alpha (ia -b) \sigma^{2}\psi^{*}_{L}=0
\end{eqnarray}
such that this equation is consistent with another equation (20) only when
\begin{eqnarray}
i \sigma^{\mu} \partial_{\mu} \psi_{L}=0,
\end{eqnarray}
or
\begin{eqnarray}
(ia -b) \sigma^{2}\psi^{*}_{L}=0
\end{eqnarray}
is satisfied. The first case (28) corresponds to our case of the
anti-commuting spinor. The
second one (29) corresponds to the commuting spinor case, which will be
shown shortly. Furthermore, for the case of
$|\alpha| \neq 1$ the Lagrangian becomes
\begin{eqnarray}
{\cal L}_{T} =-(|\alpha|^{2}-1)\left[\frac{i}{2}\psi^{\dagger}_{L}
\sigma^{\mu}\stackrel{\leftrightarrow}{\partial}_{\mu} \psi_{L}
+\frac{f}{2} \psi^{\dagger}_{L}\sigma^{2}\psi^{*}_{L}+\frac{f^{*}}{2}
\widetilde{\psi_{L}}\sigma^{2} \psi_{L} \right]
\end{eqnarray}
with
\begin{eqnarray}
f=\frac{2\alpha (ia+b)}{|\alpha|^{2}-1},
\end{eqnarray}
but the spinor $\psi_{L}$, now, has the mass squared $m^{2}$ as follows
\begin{eqnarray}
m^{2}=|f|^{2}=\frac{4 |\alpha|^{2}(a^{2}+b^{2})}{(|\alpha|^{2}-1)^{2}},
\end{eqnarray}
which is not the same as Eq. (3), that of in the original Lagrangian (12) for
any $\alpha$ with $|\alpha|
\neq 1$ such that in this case the handedness reduction is also failed. Hence,
we can conclude that there are no consistent handedness reductions for any
$\alpha$ for the case of the anti-commuting fields reproducing the result,
which is drawn from
the equations of motion in this case.

Secondly, let us consider the case of the commuting fields \footnote[8]{
Since the usual spin-statistics connections may not be applied to the tachyon
case, this would-be wrong statistics may not be ruled out from the start
although we can not obtain an affirmative answer for the statistics of the
tachyon in our problem.}, i.e.,
\begin{eqnarray}
\psi^{\dagger}_{L}(x)\psi_{L}(y)&=&\psi_{L}(y)\psi^{\dagger}_{L}(x),
 \nonumber\\
\psi_{L}(x)\psi_{L}(y)&=&\psi_{L}(y) \psi_{L}(x).
\end{eqnarray}
Then, the Lagrangian (23) reduces
\begin{eqnarray}
{\cal L}_{T}=\frac{i}{2}(|\alpha|^{2}+1)
\psi^{\dagger}_{L}\sigma^{\mu}\stackrel{\leftrightarrow}{\partial}_{\mu}
\psi_{L}.
\end{eqnarray}
But, the spinor $\psi_{L}$ in this action has different mass from that
of the original both-handed Lagrangian (12), i.e., zero-mass for any $\alpha$.
Hence,
the handedness reduction is not also consistent for any $\alpha$ in this case.
This reproduces the result drawn from the equations of motion for the
case of the commuting fields. In this way we have shown that the single
handedness
spinorial tachyon Lagrangian can not be isolated from the both handedness
spinorial tachyon
Lagrangian for the both anti-commuting and commuting fields, which directly
implying the
non-existence of the single handedness spinorial tachyon Lagrangian due to the
same reasoning in the analysis of the equations of motion.

Although our main concern is about the tachyons, it will be instructive to
compare our results with the well-known case of the bradyonic particles
allowing
the single-handed particles. To this end, we first note that the wave equation
for the general wave equation for the bradyon [3, 6]$^{2,4}$ can be
written as
\begin{eqnarray}
\left(i \gamma^{\mu}\partial_{\mu} -\lambda_{B}\right) \psi (x) =0,
\end{eqnarray}
where $\lambda_{B}=(c I +id \gamma^{5} )$, and
which has the mass squared $m^{2}$ for the spinor $\psi$ as
\begin{eqnarray}
m^{2}=(\gamma_{0} \lambda_{B})^{2} =c^{2} +d^{2} ~~>~~0
\end{eqnarray}
implying the bradyonic movement. Hence, the wave equation for the chiral
representation is written as
\begin{eqnarray}
i \overline{\sigma}^{\mu} \partial_{\mu} \psi_{R} -(c +id) \psi_{L}~=~0, \\
i \sigma^{\mu} \partial_{\mu} \psi_{L} -(c -id) \psi_{R}~=~0,
\end{eqnarray}
where both $\psi_{L}$ and $\psi_{R}$ have the same mass as $\psi$ in Eq. (36).

Now, if we try the handedness reduction by Eq. (15) as in the case of the
tachyon, then Eqs. (37) and (38) become
\begin{eqnarray}
-i \alpha \overline{\sigma}^{\mu} \sigma^{2} \partial_{\mu} \psi^{*}_{L}
        -(c +id) \psi_{L}~=~0, \\
i \sigma^{\mu} \partial_{\mu} \psi_{L}
        +\alpha(c-id)\sigma^{2} \psi^{*}_{L}~=~0.
\end{eqnarray}
These two equations would be the complex conjugations of each other if the
reduction
is consistent. To investigate this, we apply the complex conjugation to
Eq. (39), and use the Pauli's
matrices identity as in the tachyon case. Then, we obtain
\begin{eqnarray}
i \sigma^{\mu} \partial_{\mu} \psi_{L}
        +\frac{(c-id)}{\alpha^{*}}\sigma^{2} \psi^{*}_{L}~=~0,
\end{eqnarray}
which should be the same as Eq. (40) for consistency.
But, this equation is equal to Eq. (40), for non-zero $c$ or $d$ in order not
to discuss the trivially satisfying case of the luxons, only if
\begin{eqnarray}
\alpha \alpha^{*} =1
\end{eqnarray}
is satisfied \footnote[9]{
For the case $\alpha=1$, i.e., $\psi_{R}=-\sigma^{2} \psi^{*}_{L}$, the
four-component spinor $\psi$ is
usually named as the Majorana spinor, and in this case the spinor $\psi$ of
the representation
(9) is self charge-conjugate, i.e., $\psi_{c}=\psi$ under the usual charge
conjugation
\begin{eqnarray*}
\psi_{c} =\left( \begin{array}{c}
             \sigma^{2} \psi^{*}_{R} \\
             -\sigma^{2} \psi^{*}_{L} \\
            \end{array} \right)
\end{eqnarray*}
implying (electric) charge neutral. The other cases of (42)
are also charge neutral although the spinors are not self charge-conjugate.
Hence, for the case of the
bradyons, any reduced spinors (17) or (18) are charge neutral.}. This result is
derived without restricting to
any exchange algebras of the fields.

Now, let us examine how this condition is derived from the Lagrangian analysis.
To this end, we note that the Lagrangian corresponding to the equation of
motion (35) is
\begin{eqnarray}
 {\cal{L}}_{B}=\frac{i}{2} \bar{\psi}  \gamma^{\mu}
 \stackrel{\leftrightarrow}{\partial}_{\mu} \psi-  \bar{\psi}
 (c +id \gamma^{5}) \psi,
\end{eqnarray}
and hence the chiral form is
\begin{eqnarray}
{\cal L}_{B} =\frac{i}{2} \psi^{\dagger}_{L}\sigma^{\mu}
    \stackrel{\leftrightarrow}{\partial}_{\mu} \psi_{L}
            + \frac{i}{2} \psi^{\dagger}_{R}\overline{\sigma}^{\mu}
            \stackrel{\leftrightarrow}{\partial}_{\mu} \psi_{R}
              +(-c+id) \psi^{\dagger}_{L} \psi_{R}
              +(-c-id) \psi^{\dagger}_{R} \psi_{L}.
\end{eqnarray}

Now, in order to reduce this both handedness Lagrangian to single
handed (here left-handed) Lagrangian, let us replace $\psi_{R}$ with $\psi_{L}$
by Eq. (15). Then, the Lagrangian (44) reduces to
\begin{eqnarray}
{\cal L}_{B}=\frac{i}{2}\psi^{\dagger}_{L}\sigma^{\mu}
 \stackrel{\leftrightarrow}{\partial}_{\mu}
\psi_{L}
+\frac{i}{2}|\alpha|^{2} \widetilde{\psi_{L}} \widetilde{\sigma^{\mu}}
     \stackrel{\leftrightarrow}{\partial}_{\mu}
 \psi^{*}_{L}-\alpha(-c+id) \psi^{\dagger}_{L} \sigma^{2}
 \psi^{*}_{L} -\alpha^{*} (-c-id)\widetilde{\psi_{L}}\sigma^{2} \psi_{L}
\end{eqnarray}
without restricting any exchange algebras of the fields. As in parallel with
the case of the
tachyons let us consider explicitly the anti-commuting and commuting fields.
So we first consider the case of the anti-commuting fields such that
the algebra (24) is satisfied. In this case, the Lagrangian (45)
reduces to
\begin{eqnarray}
{\cal L}_{B} =(|\alpha|^{2}+1)\left[\frac{i}{2} \psi^{\dagger}_{L}\sigma^{\mu}
 \stackrel{\leftrightarrow}{\partial}_{\mu}  \psi_{L}
 +\frac{g}{2} \psi^{\dagger}_{L}\sigma^{2}
 \psi^{*}_{L}+\frac{g^{*}}{2} \widetilde{\psi_{L}}\sigma^{2} \psi_{L} \right]
\end{eqnarray}
with
\begin{eqnarray*}
g=\frac{2\alpha (-c+id)}{|\alpha|^{2}+1},
\end{eqnarray*}
but the spinor $\psi_{L}$ has the mass squared $m^{2}$ as
\begin{eqnarray}
m^{2}=|g|^{2}=\frac{4 |\alpha|^{2}(c^{2}+d^{2})}{(|\alpha|^{2}+1)^{2}},
\end{eqnarray}
which is not the same as Eq. (36), that of in the original both-handed
Lagrangian (45)
unless $|\alpha |=1$.
Hence, the handedness reduction method, which by definition should not
change any physical
properties but pick up only one-handed part, is consistent only for
$|\alpha|=1$, which reproduces the condition (42) for the anti-commuting
fields.

Secondly, if we consider the case of the commuting fields such that
algebra (33) is satisfied
, the Lagrangian (45) reduces to
\begin{eqnarray}
{\cal L}_{B}=\frac{i}{2}(1-|\alpha|^{2})
\psi^{\dagger}_{L}\sigma^{\mu}\stackrel{\leftrightarrow}{\partial}_{\mu}
\psi_{L}.
\end{eqnarray}
But, the spinor $\psi_{L}$ in this action has different mass, i.e.,
zero-mass as that of the original Lagrangian (45) unless $|\alpha |=1$ such
that the handedness reduction is not
consistent unless $|\alpha|=1$ due to the same reason as in the previous
fermionic case. This is a reproduction of the condition (42) for the commuting
fields. However, for the case of $|\alpha|=1$, the Lagrangian becomes
vanishing such that the commuting spinor may be excluded in this sense. This
can
be considered as the spin-statistics connection at the classical level. So we
have found that the handedness problem of the spinorial bradyons is drastically
different from that of the tachyons: the single-handed particles are allowed
for the bradyons but not for the tachyons.

Now, finally we will show that these results are consistent with those of the
theory without considering the handedness reduction, i.e., by considering the
single handedness two-component theory from the start. To this end, we first
note that the
most general form of the single-handed (here left-handed) and
Lorentz-invariant two-component spinor Lagrangian can be expressed as [6]
\begin{eqnarray}
{\cal L}=\frac{i}{2}\psi^{\dagger}_{L}
\sigma^{\mu}\stackrel{\leftrightarrow}{\partial}_{\mu}
\psi_{L}
+\frac{h}{2} \psi^{\dagger}_{L}\sigma^{2}\psi^{*}_{L}+\frac{h^{*}}{2}
\widetilde{\psi_{L}}\sigma^{2} \psi_{L} ,
\end{eqnarray}
which is exactly the same as the Lagrangian (30), the left handedness
Lagrangian
reduced from the both-handed original Lagrangian (23) by (15) or from (6)
by (17) except
the normalization factor and produces the wave equations
\begin{eqnarray}
i\sigma^{\mu} \partial_{\mu} \psi_{L} + h \sigma^{2} \psi^{*}_{L} =0,
\end{eqnarray}
and
\begin{eqnarray}
i\overline{\sigma}^{\mu}\sigma^{2} \partial_{\mu} \psi^{*}_{L}
+ h^{*} \sigma^{2} \psi_{L} =0
\end{eqnarray}
by varying the Lagrangian (49) with respects to $\psi^{*}_{L}$
and $\psi_{L}$ respectively. Note that these two equations are
consistent as the complex conjugated one with each others. Now, if we
calculate the mass of the spinor $\psi_{L}$ by comparing the Klein-Gordon
equation, we find that as Eq. (32)
\begin{eqnarray}
m^{2}=|h|^{2} ~\geq~0
\end{eqnarray}
such that it is easily concluded that the single-handed spinorial tachyon is
impossible although the corresponding bradyon is possible one in general.

In conclusion, we have shown in this letter that the single-handed
spinorial tachyons can not exist for any exchange algebras of the spinor fields
in
three different approaches, i.e., a) by proving the non-existence of
the reduction of the single-handed two-component tachyon from the both-handed
spinorial four-component tachyon without restricting any exchange algebras
of the fields, b) by proving the
non-existence of the previous reduction in the Lagrangian with the
specific exchange algebras of the spinors, i.e., anti-commuting and commuting
spinors which will be corresponded to the fermion and boson statistics after
the second quantization, respectively, and c) by
directly proving the non-existence of the tachyonic mode for the most
general form of the single-handed two-component spinorial spinor Lagrangian.
Hence, we conclude that at least the
both handedness are required in order that the spinorial tachyon may exist.

Now, let us consider the applicability of our result to the neutrinos
whose tachyonic property is still controversial. Since it is strongly believed
that,
if a neutrino has the tachyonic property, this should be described by the
Dirac-like equation (1) in order to have the consistency with the traditional
Dirac-equation treatment of the massless neutrinos,
the
tachyonicity of the neutrinos is governed by our result. However, there is
still no
reliable report on the existence of the right-handed neutrinos. Hence, in this
situation as far as there are no right-handed neutrinos, we conclude that our
observing left-handed neutrinos can not be the tachyons. In this way the
proposal of
the tachyonic neutrino is ruled out with the same accuracy of the
non-existence of the right-handed neutrino. However, if we really can confirm
the tachyonicity of several neutrinos in the future, this will give an
affirmative clue
to the existence of the right-handed ones for that neutrinos. \footnote[10]{
If we might find the single-handed spinorial tachyons without the
right-handed partners, we must admit the necessacity of the hypothetical number
of Footnote 6 in order to describe consistently our nature, especially the
tachyons.
This situation may be compared to the case of the Schr\"{o}dinger wave equation
of the non-relativistic quantum theory,
where the introduction of the pure-imaginary number is inevitable.}
Of course, the reverse reasoning needs not be true.

Finally, we comment that the widely used mechanism for the smallness of
the neutrino, see-saw mechanism [7], which needs two handedness of the
neutrinos, may not exclude the possibility of the tachyonicity.
\\

The present work was supported by the Basic Science Research Institute
program, Ministry of Education, Project No. BSRI-95-2414.

\newpage
\begin{center}
{\large \bf REFERENCES}
\end{center}
\begin{description}
\item{[1]} R. G. H. Robertson, in: 14th
International Workshop on Weak Interactions and Neutrino (Seoul, July 1993),
eds. J. E. Kim and S. K. Kim (World Scientific, Singapore, 1994) and references
therein; Particle Data
Group, Phys. Rev. D 50 (1994) 1390; A. Chodos and V. A. Kosteleck\'{y},
Phys. Lett. B 336 (1994) 295 and references therein.

\item{[2]}
S. Tanaka, Prog. Theor. Phys. 24 (1960) 171; O. M. P. Bilaniuk,
V. K. Deshapande, and E. C. G. Sudarshan, Am. J. Phys. 30 (1962)
718; G. Feinberg, Phys. Rev. D 159 (1967) 1089.

\item{[3]}
S. Tanaka, quoted in Ref. [2]; J. Bandukwala
and D. Shay, Phys. Rev. D 9 (1974) 889; A. Chodos, A. I. Houser, and
V. A. Kosteleck\'{y}, Phys. Lett. B 150 (1985) 431; E. Giametto et al.,
Phys. Lett. B 178 (1986) 115; A. Chodos
et al., Mod. Phys. Lett. A 7 (1992) 467,
L. C. Biedenharn, Phys. Lett B 158 (1985) 227; R. J. Hughes and G. J.
Stephenson Jr., Phys. Lett. B 244 (1990) 95.

\item{[4]} J. D. Bjorken and S. D. Drell, Relativistic Quantum Fields
(McGraw-Hill, Inc., 1965); C. Itzykson and J. B. Zuber, Quantum Field Theory
(McGraw-Hill, Inc., 1980).

\item{[5]} S. Tanaka, quoted in Ref. [2]; M. E. Arons and E. C. G. Sudarshan,
Phys. Rev. 173 (1968) 1622; J. Dhar and E. C. G. Sudarshan, {\it ibid}.
174 (1968) 1808; G. Ecker, Ann. Phys. 58 (1970) 303;
B. Schroer, Phys. Rev. D 3 (1971) 1964; K. Kamoi and S. Kamefuchi,
Prog. Theor. Phys. 45 (1971) 1946.

\item{[6]} P. Ramond, Field Theory: A Modern Primer (Addison-Wesley Publishing
Company, Inc., 1989); L. S. Brown, Quantum Field Theory (Cambridge University
Press, 1992).

\item{[7]} M. Gell-Mann, P. Ramond, and R. Slansky, in: Supergravity, eds.
P. van Nieuwenhuizen and D. Freeman (North Holland, Amsterdam, 1979);
T. Yanagida, Prog. Theor. Phys. B 135 (1978) 66;
R. N. Mohapatra, P. B. Pal, Massive Neutrinos in Physics and Astrophysics
(World Scientific, London, 1991).

\end{description}
\end{document}